\documentclass[%
 reprint,
 letter,
 superscriptaddress,
 showpacs,preprintnumbers,
 amsmath,amssymb,
 aps,
 prl
]{revtex4-1}

\usepackage{graphicx}
\usepackage{booktabs,dcolumn}

\usepackage[draft]{pgf}

\begin{document}

\title{First Accurate Normalization of the 
$\boldsymbol{\beta}$-delayed $\boldsymbol{\alpha}$ Decay of $\boldsymbol{^{16}}$N\\ 
and Implications 
for the $\boldsymbol{^{12}}\text{C}\boldsymbol{(\alpha,\gamma)^{16}}\text{O}$ 
Astrophysical Reaction Rate}


\author{O.~S.~Kirsebom}
\email[Corresponding author: ]{oliskir@phys.au.dk}
\affiliation{Department of Physics and Astronomy, Aarhus University, DK-8000 Aarhus C, Denmark}


\author{O.~Tengblad}
\affiliation{Instituto de Estructura de la Materia, CSIC, E-28006 Madrid, Spain}

\author{R.~Lica}
\affiliation{CERN, CH-1211 Geneva 23, Switzerland}
\affiliation{Horia Hulubei National Institute for Physics and Nuclear Engineering (IFIN-HH), RO-077125 Bucharest-Magurele, Romania}

\author{M.~Munch}
\affiliation{Department of Physics and Astronomy, Aarhus University, DK-8000 Aarhus C, Denmark}

\author{K.~Riisager}
\affiliation{Department of Physics and Astronomy, Aarhus University, DK-8000 Aarhus C, Denmark}

\author{H.~O.~U.~Fynbo}
\affiliation{Department of Physics and Astronomy, Aarhus University, DK-8000 Aarhus C, Denmark}

\author{M.~J.~G.~Borge}
\affiliation{Instituto de Estructura de la Materia, CSIC, E-28006 Madrid, Spain}
\affiliation{CERN, CH-1211 Geneva 23, Switzerland}

\author{M.~Madurga}
\affiliation{CERN, CH-1211 Geneva 23, Switzerland}

\author{I.~Marroquin}
\affiliation{Instituto de Estructura de la Materia, CSIC, E-28006 Madrid, Spain}


\author{A.~N.~Andreyev}
\affiliation{Department of Physics, University of York, York YO10 5DD, United Kingdom}

\author{T.~A.~Berry}
\affiliation{Department of Physics, University of Surrey, Guildford, GU2 7XH, United Kingdom}

\author{E.~R.~Christensen}
\affiliation{Department of Physics and Astronomy, Aarhus University, DK-8000 Aarhus C, Denmark}

\author{P.~D{\'i}az~Fern{\'a}ndez}
\affiliation{Department of Physics, Chalmers University of Technology, S-41296 G{\"o}teborg, Sweden}

\author{D.~T.~Doherty}
\affiliation{Department of Physics, University of York, York YO10 5DD, United Kingdom}

\author{P.~Van Duppen}
\affiliation{KU Leuven, Instituut voor Kern- en Stralingsfysica, 3001 Leuven, Belgium}

\author{L.~M.~Fraile}
\affiliation{Grupo de F{\'i}sica Nuclear, Universidad Complutense de Madrid, E-28040 Madrid, Spain}

\author{M.~C.~Gallardo}
\affiliation{Grupo de F{\'i}sica Nuclear, Universidad Complutense de Madrid, E-28040 Madrid, Spain}

\author{P.~T.~Greenlees}
\affiliation{University of Jyvaskyla, Department of Physics, P.O.\ Box 35, FI-40014 University of Jyvaskyla, Finland}
\affiliation{Helsinki Institute of Physics, University of Helsinki, P.O.\ Box 64, FI-00014 Helsinki, Finland}

\author{L.~J.~Harkness-Brennan}
\affiliation{Oliver Lodge Laboratory, University of Liverpool, Liverpool L69 7ZE, United Kingdom}

\author{N.~Hubbard}
\affiliation{Department of Physics and Astronomy, Aarhus University, DK-8000 Aarhus C, Denmark}
\affiliation{Department of Physics, University of York, York YO10 5DD, United Kingdom}

\author{M.~Huyse}
\affiliation{KU Leuven, Instituut voor Kern- en Stralingsfysica, 3001 Leuven, Belgium}

\author{J.~H.~Jensen}
\affiliation{Department of Physics and Astronomy, Aarhus University, DK-8000 Aarhus C, Denmark}

\author{H.~Johansson}
\affiliation{Department of Physics, Chalmers University of Technology, S-41296 G{\"o}teborg, Sweden}

\author{B.~Jonson}
\affiliation{Department of Physics, Chalmers University of Technology, S-41296 G{\"o}teborg, Sweden}

\author{D.~S.~Judson}
\affiliation{Oliver Lodge Laboratory, University of Liverpool, Liverpool L69 7ZE, United Kingdom}

\author{J.~Konki}
\affiliation{CERN, CH-1211 Geneva 23, Switzerland}
\affiliation{University of Jyvaskyla, Department of Physics, P.O.\ Box 35, FI-40014 University of Jyvaskyla, Finland}
\affiliation{Helsinki Institute of Physics, University of Helsinki, P.O.\ Box 64, FI-00014 Helsinki, Finland}

\author{I.~Lazarus}
\affiliation{STFC Daresbury, Daresbury, Warrington WA4 4AD, United Kingdom}

\author{M.~V.~Lund}
\affiliation{Department of Physics and Astronomy, Aarhus University, DK-8000 Aarhus C, Denmark}

\author{N.~Marginean}
\affiliation{Horia Hulubei National Institute for Physics and Nuclear Engineering (IFIN-HH), RO-077125 Bucharest-Magurele, Romania}

\author{R.~Marginean}
\affiliation{Horia Hulubei National Institute for Physics and Nuclear Engineering (IFIN-HH), RO-077125 Bucharest-Magurele, Romania}

\author{A.~Perea}
\affiliation{Instituto de Estructura de la Materia, CSIC, E-28006 Madrid, Spain}

\author{C.~Mihai}
\affiliation{Horia Hulubei National Institute for Physics and Nuclear Engineering (IFIN-HH), RO-077125 Bucharest-Magurele, Romania}

\author{A.~Negret}
\affiliation{Horia Hulubei National Institute for Physics and Nuclear Engineering (IFIN-HH), RO-077125 Bucharest-Magurele, Romania}

\author{R.~D.~Page}
\affiliation{Oliver Lodge Laboratory, University of Liverpool, Liverpool L69 7ZE, United Kingdom}


\author{V.~Pucknell}
\affiliation{STFC Daresbury, Daresbury, Warrington WA4 4AD, United Kingdom}

\author{P.~Rahkila}
\affiliation{University of Jyvaskyla, Department of Physics, P.O.\ Box 35, FI-40014 University of Jyvaskyla, Finland}
\affiliation{Helsinki Institute of Physics, University of Helsinki, P.O.\ Box 64, FI-00014 Helsinki, Finland}


\author{O.~Sorlin}
\affiliation{CERN, CH-1211 Geneva 23, Switzerland}
\affiliation{GANIL, CEA/DSM-CNRS/IN2P3, Bvd Henri Becquerel, 14076 Caen, France}

\author{C.~Sotty}
\affiliation{Horia Hulubei National Institute for Physics and Nuclear Engineering (IFIN-HH), RO-077125 Bucharest-Magurele, Romania}

\author{J.~A.~Swartz}
\affiliation{Department of Physics and Astronomy, Aarhus University, DK-8000 Aarhus C, Denmark}

\author{H.~B.~S{\o}rensen}
\affiliation{Department of Physics and Astronomy, Aarhus University, DK-8000 Aarhus C, Denmark}

\author{H.~T{\"o}rnqvist}
\affiliation{Institut f{\"u}r Kernphysik, Technische Universit{\"a}t Darmstadt, Darmstadt, Germany}
\affiliation{GSI Helmholtzzentrum f{\"u}r Schwerionenforschung, Darmstadt, Germany}

\author{V.~Vedia}
\affiliation{Grupo de F{\'i}sica Nuclear, Universidad Complutense de Madrid, E-28040 Madrid, Spain}


\author{N.~Warr}
\affiliation{Institut f\"ur Kernphysik, Universit\"at zu K\"oln, D-50937 K\"oln, Germany}

\author{H.~De Witte}
\affiliation{KU Leuven, Instituut voor Kern- en Stralingsfysica, 3001 Leuven, Belgium}

\date{\today}

\begin{abstract}

The $^{12}\text{C}(\alpha,\gamma){}^{16}\text{O}$ 
reaction plays a central role in astrophysics, 
but its cross section at energies relevant 
for astrophysical applications is only poorly 
constrained by laboratory data. %
The reduced $\alpha$ width, $\gamma_{11}$, of the 
bound $1^-$ level in $^{16}$O is particularly important 
to determine the cross section.
The magnitude of $\gamma_{11}$ is determined via 
sub-Coulomb $\alpha$-transfer reactions or the 
$\beta$-delayed $\alpha$ decay of $^{16}$N, but 
the latter approach is presently hampered by the 
lack of sufficiently precise data on the $\beta$-decay 
branching ratios. %
Here we report improved branching ratios for the bound 
$1^-$ level [$b_{\beta,11} = (5.02\pm 0.10)\times 10^{-2}$] 
and for $\beta$-delayed $\alpha$ 
emission [$b_{\beta\alpha} = (1.59\pm 0.06)\times 10^{-5}$]. %
Our value for $b_{\beta\alpha}$ is 33\% larger than previously 
held, leading to a substantial increase in $\gamma_{11}$. %
Our revised value for $\gamma_{11}$ is in good agreement with 
the value obtained in $\alpha$-transfer studies and the weighted 
average of the two gives a robust and precise determination of 
$\gamma_{11}$, which provides significantly improved constraints 
on the $^{12}$C$(\alpha,\gamma)$ cross section in the energy range 
relevant to hydrostatic He burning.

\end{abstract}

\maketitle

In the hot and dense interior of stars,
helium is burned into carbon and oxygen 
by means of the triple-$\alpha$ reaction 
and the $^{12}\text{C}(\alpha,\gamma)$ reaction. %
The rates of the two reactions 
regulate the relative production of carbon 
and oxygen---a quantity of paramount 
importance in astrophysics affecting  
everything from grain formation in stellar winds 
to the late evolution of massive stars and 
the composition of type-Ia supernova progenitors~\cite{deBoer2017}. %
At the temperatures characteristic of hydrostatic
He burning, the triple-$\alpha$ reaction is dominated 
by a single, narrow resonance---the so-called 
Hoyle resonance---and hence it has been 
possible to constrain the reaction rate 
through measurements of the properties of 
the Hoyle resonance. %
In contrast, the $^{12}\text{C}(\alpha,\gamma)$ 
reaction receives contributions from several
levels in $^{16}$O, which, as it happens, all
lie outside the energy window where 
thermal fusion of $\alpha+{}^{12}\text{C}$ in 
the stellar environment is 
efficient---the so-called Gamow window. 
This makes the task of determining the 
$^{12}\text{C}(\alpha,\gamma)$ rate rather complex. %
While the triple-$\alpha$ rate is now
considered known within 10\% in the energy range 
relevant to hydrostatic He burning~\cite{fynbo2014}, 
with efforts underway to reduce the uncertainty 
to 5\%~\cite{tur2008, kibedi2012}, the uncertainty 
on the $^{12}\text{C}(\alpha,\gamma)$ 
rate was recently estimated 
to be at least 20\% which is insufficient for 
several astrophysical applications~\cite{deBoer2017}. %

The $^{12}\text{C}(\alpha,\gamma)$ cross section 
has been measured down to center-of-mass energies of 
$\approx 1.0$~MeV, but the rapidly 
decreasing tunneling probability makes it 
challenging to extend the measurements to lower 
energies and practically impossible to reach the 
Gamow energy of 0.3~MeV. %
According to current understanding~\cite{deBoer2017}, 
the capture cross section at 0.3~MeV receives 
its largest single contribution from the high-energy 
tail of the bound $1^{-}$ level in $^{16}$O, situated 
at an excitation energy of $E_x=7.12$~MeV only 45~keV 
below the $\alpha+{}^{12}\text{C}$ threshold 
. %
The reduced $\alpha$ width of this 
level, $\gamma_{11}$, provides a measure 
of how strongly the level couples to the
$\alpha+{}^{12}\text{C}$ channel. %
Therefore, $\gamma_{11}$ is a critical quantity 
in determining the level's contribution to the 
capture cross section at 0.3~MeV and, more 
generally, in constraining the extrapolation of the 
$^{12}\text{C}(\alpha, \gamma)$ cross section to the 
energy range relevant for stellar helium burning. %
Specifically, the dominant term in the expression 
for the $E1$ capture cross section (see, e.g., Eq.~(6) 
in Ref.~\cite{azuma1994}) is proportional to 
$P_1\gamma_{11}^2$ where $P_1$ 
is the $p$-wave penetration factor of the $\alpha+{}^{12}\text{C}$ 
channel. %

The magnitude of $\gamma_{11}$ can be determined 
from the $\beta$-delayed $\alpha$ spectrum 
($\beta\alpha$ spectrum) of $^{16}$N~\cite{barker1971}, 
but currently this approach is hindered by uncertainties in the normalization 
of the spectrum~\cite{humblet1991, buchmann2009} as the inferred 
value for $\gamma_{11}$ is strongly correlated with the assumed 
$\beta$-decay branching ratios ($\gamma_{11}^2 \propto 
b_{\beta\alpha} / b_{\beta,11}$, see Supplemental Material).
Furthermore, the spectral form is not well determined experimentally 
due to small but significant discrepancies between existing measurements. %
Here, we focus our attention on the two high-precision 
spectra of Refs.~\cite{azuma1994, tang2010} while disregarding a 
handful of other spectra, including those of Refs.~\cite{hattig1970, france2007}, 
which all ``retain significant experimental effects''~\cite{deBoer2017}. %

In this Letter, we report on an experimental 
study of the $\beta\alpha$ decay of $^{16}$N 
in which the unique radioactive-isotope 
production capabilities of the ISOLDE facility~\cite{catherall2017} 
are exploited to provide the first accurate 
and precise determination of $b_{\beta\alpha}$. 
We also present a novel $R$-matrix analysis of 
the $\beta\alpha$ spectra of Refs.~\cite{azuma1994, tang2010}, 
propose a resolution to the discrepancies 
between the two spectra,
and extract an improved value 
for $P_1\gamma_{11}^2$ which is in 
good agreement with the value 
inferred from sub-Coulomb $\alpha$-transfer 
reactions. %
Finally, we comment on the implications of 
our findings for the determination of the 
$^{12}\text{C}(\alpha,\gamma)$ cross section 
at 0.3~MeV. %
A detailed account of the experimental work 
and the $R$-matrix analysis will be published 
separately~\cite{kirsebom_in_prep_2018}.

The experiment was performed at the 
ISOLDE radioactive-beam facility of CERN~\cite{catherall2017}. 
Radioactive isotopes were produced by the impact 
of a 1.4-GeV proton beam on a nano-structured CaO 
target~\cite{ramos2014}, before being ionized 
in a cooled plasma ion source and accelerated 
through an electrostatic potential difference 
of 30~kV. %
Ions with the desired mass-to-charge ($A/q$) 
ratio were selected in the High-Resolution 
Separator and guided to the ISOLDE Decay 
Station~\cite{fynbo2017} where their 
decay was studied. %
The ions were stopped in a thin ($33\pm3$~$\mu$g/cm$^2$) 
carbon foil surrounded by five double-sided silicon 
strip detectors (DSSD) and four high-purity germanium 
(HPGe) clovers, allowing for the simultaneous 
detection of charged particles and $\gamma$ rays. %
Meanwhile, auxiliary detectors were used to check that 
the beam was being fully transmitted to the center 
of the setup and stopped in the foil. %
During five days of data taking, 
the $\beta\alpha$ decay of $^{16}$N was studied 
mainly on $A/q=30$ ($^{16}$N$^{14}$N$^+$) but 
also on $A/q=31$ ($^{16}$N$^{14}$N$^{1}$H$^+$). %
%
Additionally, the decays of $^{17}$Ne ($\beta\gamma$, 
$\beta p$, $\beta\alpha$), $^{18}$N ($\beta\gamma$, 
$\beta\alpha$), and $^{34}$Ar ($\beta\gamma$) were studied 
on $A/q=17$, 32, and 34, providing crucial
data for the efficiency calibration of the HPGe array 
and the energy calibration of the DSSD array.%

\begin{figure}
    \includegraphics[width=0.99\columnwidth,clip=true,trim=0 5 0 5]{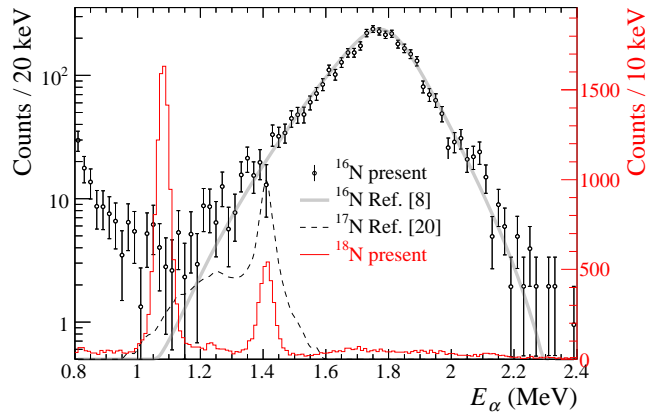}
    \caption{\label{fig:alpha} $\beta$-delayed $\alpha$ spectra obtained in one of the 60-$\mu$m thick DSSDs on $A/q=30$ (black circles) and $32$ (red histogram). %
    The two narrow $\alpha$ lines from the $\beta\alpha$ decay of $^{18}$N feature prominently in the spectrum obtained on $A/q=32$, 
    while the spectrum obtained on $A/q=30$ is due 
    almost entirely to the $\beta\alpha$ decay of $^{16}$N except for a $(2.0\pm 0.4)\%$ contamination from the $\beta\alpha$ decay of $^{17}$N (dashed curve) which has been subtracted. The $R$-matrix fit to the $^{16}$N spectrum of Ref.~\cite{azuma1994} (downscaled and properly corrected for experimental resolution) is also shown (thick, gray curve).}
\end{figure}

Three of the DSSDs were sufficiently thin 
(40~$\mu$m and 60~$\mu$m) to allow the $\alpha$ spectrum of 
$^{16}$N to be clearly separated from the $\beta$ background. %
The other two DSSDs were much thicker (300~$\mu$m and 
1~mm) and served primarily to detect the $\beta$ 
particles. %
The distortions of the $\alpha$ spectrum due to $\beta$ 
summing was negligible due to the high granularity of the 
DSSDs~\cite{kirsebom2011}. %
Fig.~\ref{fig:alpha} shows 
the $\alpha$ spectrum obtained in one of the thin 
DSSDs on $A/q=30$ during 32 hours of measurement 
at an average $^{16}$N implantation rate of 
$2\times 10^{4}$~ions/s. %
The two narrow peaks at $E_{\alpha} = 1081\pm 1$ 
and $1409\pm 1$~keV in the $\beta\alpha$ 
spectrum of $^{18}\text{N}$~\cite{hahn1993, tunl18} 
obtained on $A/q=32$ were used to determine 
the detector response and energy calibration. %
The energy resolution was 30~keV (FWHM) for the 
two 60-$\mu$m DSSDs and 70~keV for the 40-$\mu$m DSSD. %

\begin{figure}
    \includegraphics[width=0.99\columnwidth,clip=true,trim=0 52 0 0]{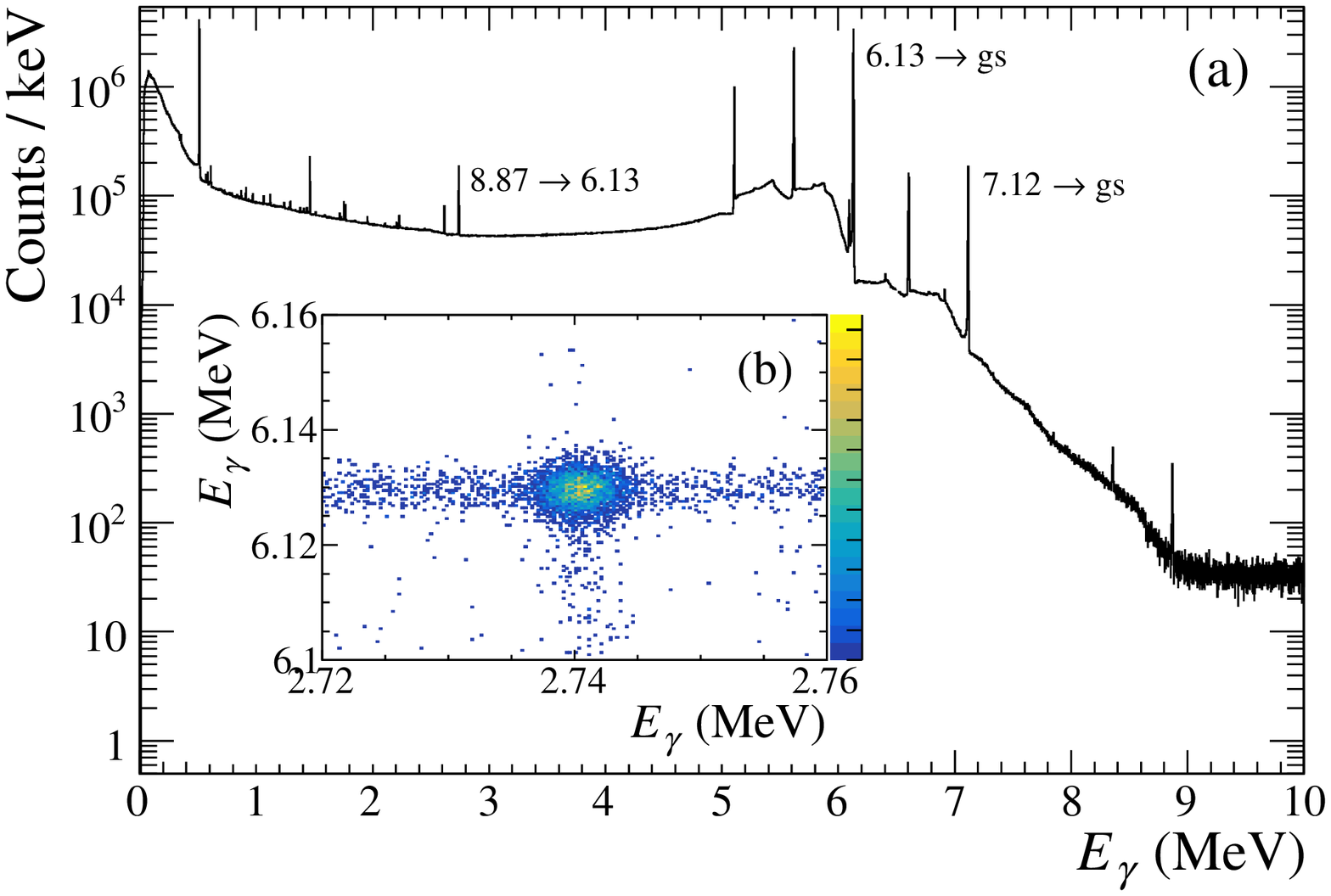}  
    \includegraphics[width=0.99\columnwidth,clip=true,trim=0 0 0 20]{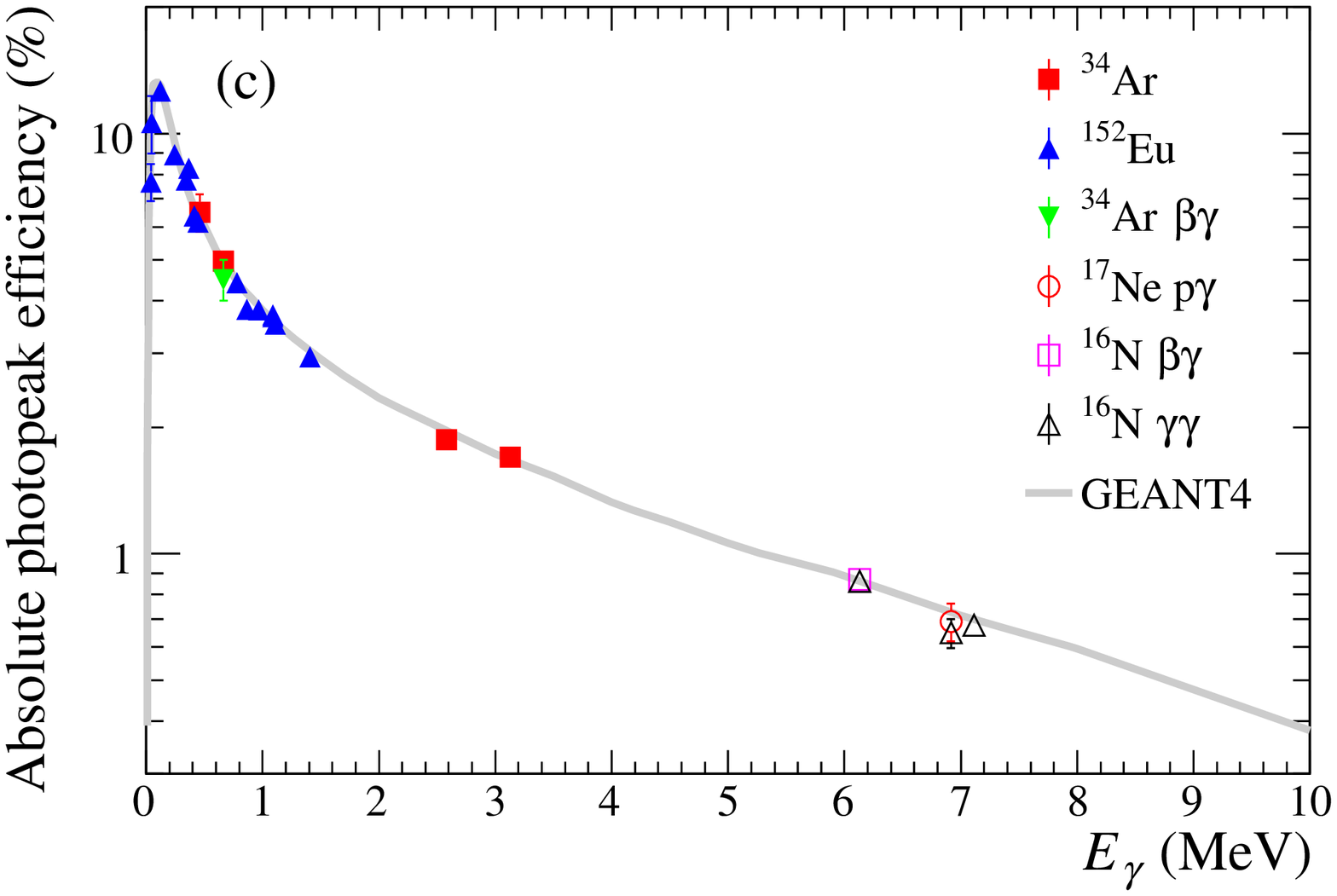}  
    \caption{\label{fig:gamma} (a)~$\gamma$-ray spectrum from the $\beta$ decay of $^{16}$N with main 
    transitions indicated. (b)~$\gamma\gamma$ coincidence spectrum
    zoomed in on the $8.87\rightarrow 6.13\rightarrow\text{g.s.}$ cascade. 
    (c)~Experimentally determined 
    and simulated $\gamma$-ray detection efficiency.}
\end{figure}

The top panel of Fig.~\ref{fig:gamma} shows the $\gamma$-ray spectrum 
measured in the HPGe clovers. The spectrum exhibits the characteristic $\gamma$ rays 
from the decay of $^{16}$N~\cite{tunl16}, most notably the prominent 
lines at 2.74, 6.13, and 7.12~MeV. %
Additionally, the spectrum provides evidence for only one other 
$\beta$-delayed particle emitter, namely, $^{17}$N, present at a 
level of 1.3\% relative to $^{16}$N, as inferred 
from the observation of its 0.871-MeV and 2.18-MeV 
$\gamma$ rays. %
Based on the known $\beta\alpha$ 
branching ratio of $^{17}$N of 
$(2.5\pm 0.4)\times 10^{-5}$~\cite{dombsky1994}, we determine the 
level of $^{17}$N contamination in our 
$\alpha$ spectrum to be $(2.0\pm 0.4)\%$. %
In order to convert the observed $\gamma$-ray yields 
to intensity ratios it is necessary to correct for 
the energy dependent detection efficiency of the 
HPGe array. %
An absolutely calibrated $^{152}$Eu source was used to
determine the detection efficiency at low energies, 
while $\beta\gamma$, 
$\gamma\gamma$, and $p\gamma$ coincidence-data were 
used to extend the efficiency calibration to higher 
energies. %
A GEANT4 simulation~\cite{sotty_IDS_GEANT4_in_preparation}, normalized only to the 
$^{152}$Eu data, was used to validate the efficiency 
calibration. As seen in Fig.~\ref{fig:gamma}~(c), 
there is excellent agreement across the entire energy 
range. %
Particular attention was paid to the 6.13-MeV $\gamma$ ray 
since it is used for the overall normalization. %
Using the $\gamma\gamma$ coincidences due to the $8.87\rightarrow 6.13\rightarrow\text{g.s.}$ 
cascade (Fig.~\ref{fig:gamma}~(b)) and $\beta\gamma$ coincidences, 
the detection efficiency at 6.13~MeV was determined with a precision of 1.4\%. %
After correcting the observed $\gamma\gamma$ coincidence yield for the 
known angular correlation~\cite{vermeer1982}, the two approaches 
($\gamma\gamma$ and $\beta\gamma$) gave fully consistent results. %
%
%
%

Based on the relative $\gamma$-ray yields, we 
determine the $\beta$-decay branching ratio to the 7.12-MeV level 
in $^{16}$O to be $b_{\beta,11} = (5.02\pm 0.10)\times 10^{-2}$ 
in agreement with Refs.~\cite{tunl16, millar1951, toppel1956, alburger1959, tang2010}, 
but with a reduced uncertainty due to the precise 
efficiency calibration and high energy resolution of the present study. %
Based on the number of detected $\alpha$ particles, 
the measured 6.13-MeV $\gamma$-ray yield, and the known 
relative intensity of the 6.13-MeV $\gamma$-ray 
line ($0.670\pm0.006$~\cite{tunl16, warburton1984, heath1985}), we determine the 
branching ratio for $\alpha$ emission to be $b_{\beta\alpha} 
= (1.59\pm 0.06)\times 10^{-5}$ with the following error budget: %
$\alpha$-particle detection efficiency, 3.0\%; $\gamma$-ray detection 
efficiency, 1.4\%; $\alpha$-particle counting uncertainty, 1.3\%; tabulated 
intensity of the 6.13-MeV $\gamma$ ray, 0.9\%; and subtraction of the 
$^{17}$N contamination, 0.4\%. %
When added in quadrature these uncertainties combine to give 
the quoted total uncertainty of 3.8\% on $b_{\beta\alpha}$. %
Our value for $b_{\beta\alpha}$ is significantly 
larger than the literature value of 
$(1.20\pm0.05)\times10^{-5}$~\cite{tunl16, kaufmann1961}, but
consistent with the less precise values of 
$(1.3\pm 0.3)\times 10^{-5}$ obtained by Ref.~\cite{zhao1993} 
and $(1.49\pm0.05\text{(stat)}^{+0.0}_{-0.10}\text{(sys)})\times10^{-5}$ 
obtained by us in a previous study using a 
different experimental technique~\cite{refsgaard2016}. %

In order to parametrize the shape of the 
$\alpha$ spectrum, we adopt an $R$-matrix 
model similar to that of Refs.~\cite{azuma1994, tang2010}, %
consisting of two physical $p$-wave  
levels at $E_x = 7.12$ and 9.59~MeV, two physical $f$-wave 
levels at $E_x = 6.13$ and 11.60~MeV, and 
a $p$-wave background pole at higher energy. %
The $R$-matrix model of Refs.~\cite{azuma1994, tang2010} 
additionally includes an $f$-wave background 
pole with zero feeding, but we find that the 
inclusion of such a pole only gives a marginal 
improvement of $\chi^2$ and a slightly worse 
$\chi^2/N$ and hence we do not include it. 
On the other hand, we allow the feeding of the 
11.60-MeV level, which was also set to zero 
in Refs.~\cite{azuma1994, tang2010}, to vary freely. 
Our analysis differs from those 
of Refs.~\cite{azuma1994, tang2010} in 
a few significant respects: %
First and most importantly, the analyses of Refs.~\cite{azuma1994, tang2010} 
were aimed at determining the capture cross 
section at 0.3~MeV and therefore involved the 
simultaneous fitting of $\beta\alpha$-decay data, 
$\alpha$-scattering data, and $\alpha$-capture data. %
Our analysis, on the other hand, is aimed at
determining the constraints imposed on 
$\gamma_{11}$ by the $\beta\alpha$-decay data 
alone and at resolving the discrepancies 
between Refs.~\cite{azuma1994, tang2010}, 
and hence we restrict our attention to the $\beta\alpha$-decay data. %
%
%
We also adopt our improved values for 
$b_{\beta,11}$ and $b_{\beta\alpha}$, and 
we fix the asymptotic normalization 
coefficient (ANC) of the 6.13-MeV level 
to the rather precise value of 
$C=139\pm9$~fm$^{-1/2}$ inferred from 
sub-Coulomb transfer reactions \cite{avila2015}. %
All $R$-matrix calculations have been performed with 
the code ORM~\cite{ORM}. Further details 
provided in Supplemental Material. %

\begin{figure*}
    \begin{minipage}{0.49\textwidth}%
        \includegraphics[width=0.99\columnwidth,clip=true,trim=0 0 60 0]{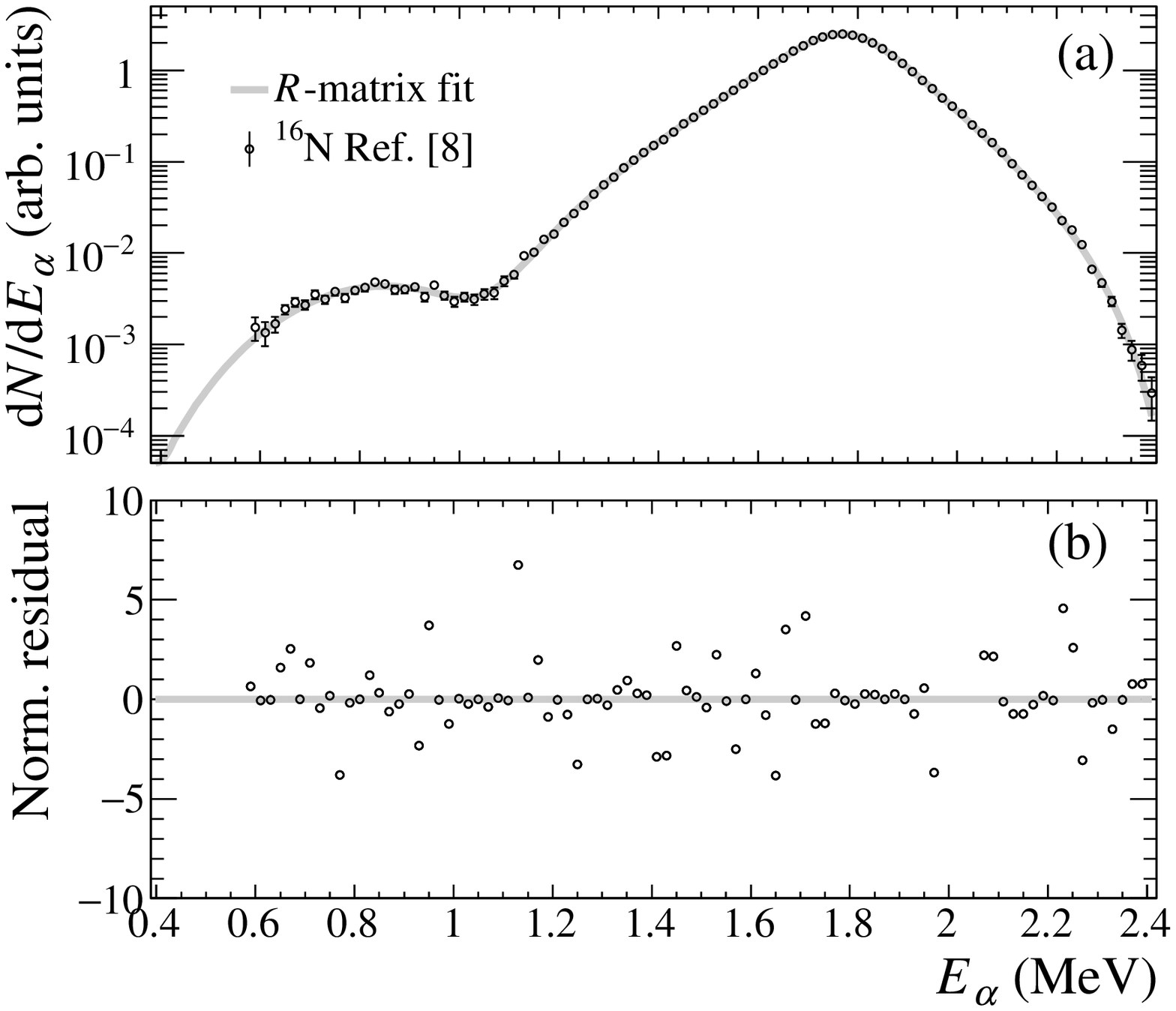}%
    \end{minipage}%
\begin{minipage}{0.49\textwidth}%
        \includegraphics[width=0.99\columnwidth,clip=true,trim=60 0 0 0]{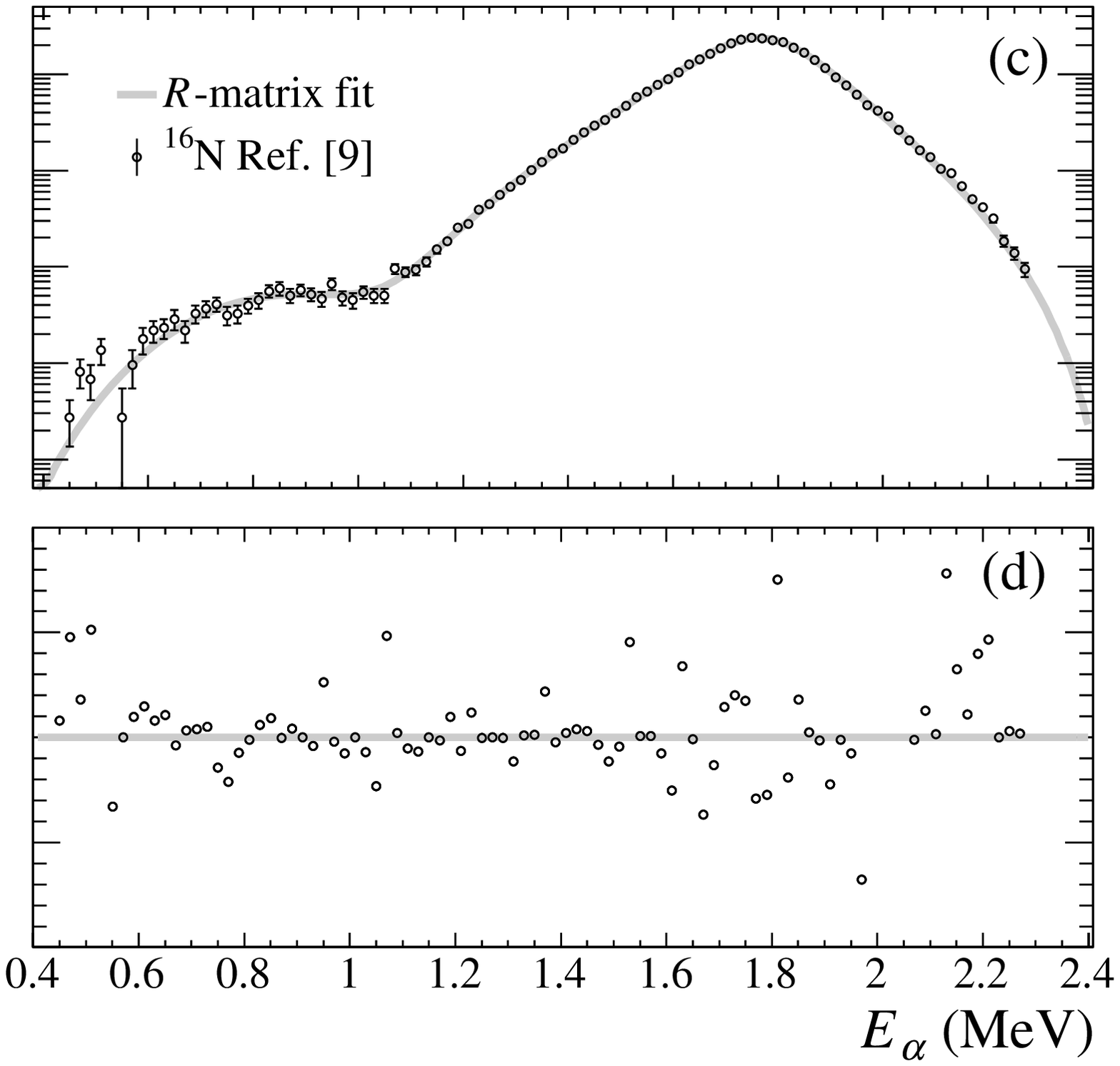}%
    \end{minipage}
    \caption{\label{fig:fit-to-Azuma-and-Tang} (a), (c):  $R$-matrix fits to the $\beta\alpha$ spectra 
        of Refs.~\cite{azuma1994, tang2010}.  
    (b), (d): Normalized residuals.}
\end{figure*}

Following Refs.~\cite{azuma1994, tang2010} we ignore 
the four data points in the vicinity of the narrow 
$2^+$ level at $E_x=9.68$~MeV. %
Allowing the channel radius to vary, 
we obtain a very good fit to the spectrum of Ref.~\cite{azuma1994} 
($\chi^2/N = 94.3/79 = 1.19$, $P_{\chi^2>94.3} = 0.116$, Fig.~\ref{fig:fit-to-Azuma-and-Tang} left panel) yielding 
\begin{equation}\label{eq:gamma}
P_1 \gamma_{11}^2 = 5.17\pm0.75\text{(stat)} \pm 0.54\text{(sys)}\; \mu\text{eV} 
\end{equation}
(with $P_1$ evaluated at 0.3~MeV) and a preferred channel radius of $6.35$~fm. %
The largest contribution to the systematic 
uncertainty comes from the energy calibration (3.8\%)
with smaller contributions from $b_{\beta\alpha}$  
(2.7\%) and $b_{\beta,11}$ (2.0\%) and 
even smaller contributions from the subtraction of
$^{17}$N and $^{18}$N impurities (1.0\%), the ANC 
of the 6.13-MeV 
level (0.4\%), and the energy resolution (0.3\%). %
Using the old branching ratio of 
$b_{\beta\alpha}=1.20\times 10^{-5}$~\cite{tunl16, kaufmann1961}, 
we obtain $P_1\gamma_{11}^2 = 3.92\pm0.57\text{(stat)}\; \mu$eV 
with no change in fit quality. Thus, our revised value 
for $b_{\beta\alpha}$ leads to a 32\% increase in $P_1\gamma_{11}^2$. %
The precise effect on the $E1$ capture cross section is 
difficult to determine since it requires a simultaneous fit to the 
$\beta\alpha$ spectrum, $\alpha$-capture data, and $\alpha$-scattering data, which is beyond the scope of the present study. An accurate estimate can, however, be obtained by adopting the best-fit parameters of Ref.~\cite{azuma1994} and only modify the value of $\gamma_{11}$. %
Doing so, one finds a 24\% increase in the $E1$ capture cross section at 0.3~MeV, implying an upward shift of the best estimate of the astrophysical $S$-factor from $S_{E1}(0.3) = 79$~keV~b~\cite{azuma1994} to $S_{E1}(0.3) = 98$~keV~b. %

\begin{figure}
    \includegraphics[width=0.99\columnwidth]{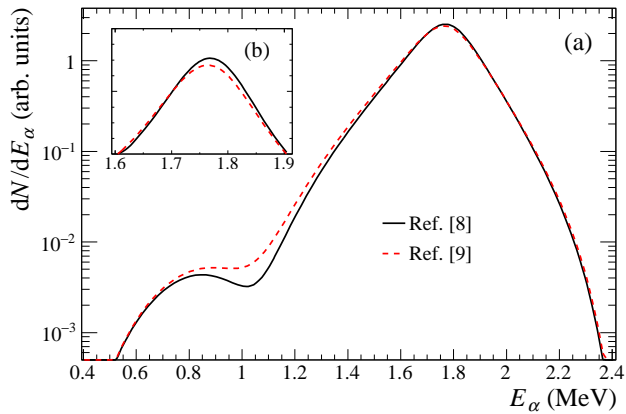}%

    \caption{\label{fig:azuma-tang} (a) Comparison of the $R$-matrix distributions determined from the
    $\beta\alpha$ spectra of Refs.~\cite{azuma1994, tang2010}. (b)~Zoom-in on the maximum of the distribution.}
\end{figure}

We are unable to obtain a satisfactory fit to the spectrum of Ref.~\cite{tang2010} 
($\chi^2/N = 114.9/79 = 1.45$, $P_{\chi^2>114.9} = 0.005$, 
Fig.~\ref{fig:fit-to-Azuma-and-Tang} right panel). %
Also, the channel radius preferred by the fit is significantly 
smaller ($5.35$~fm). %
Yet, we obtain $P_1 \gamma_{11}^2 = 6.82\pm0.65\text{(stat)}\; \mu$eV 
in fair agreement with Eq.~(\ref{eq:gamma}). 
Given the discrepancies between the two 
spectra~\cite{fn1}, it is a little surprising 
that we obtain almost agreeing values for $P_1 \gamma_{11}^2$. 
As seen in Fig.~\ref{fig:azuma-tang}, 
the dip around $E_{\alpha}=1.0$~MeV is less pronounced in 
the spectrum of Ref.~\cite{tang2010}, and the main peak is slightly 
wider and shifted by $-6$~keV relative to the spectrum of 
Ref.~\cite{azuma1994}. %
However, a detailed analysis reveals the agreement 
to be little more than a lucky coincidence: The less pronounced 
dip favours a larger $\gamma_{11}$ value, but the 
downward energy shift has the opposite effect on $\gamma_{11}$ so 
the two differences almost cancel out. %

The spectrum obtained in the present work 
contains significantly fewer counts ($1.07\times 10^4$) 
than the spectra of Refs.~\cite{azuma1994, tang2010} 
($1.03\times 10^6$ and $2.75\times 10^5$) and hence does not 
impose any useful constraints on $P_1\gamma_{11}^2$. %
Our spectrum does, however, impose useful constraints on the 
position of the maximum of the $R$-matrix distribution. %
Taking into account the uncertainty on the energy calibration, 
the maximum is found to be consistent with Ref.~\cite{azuma1994}, 
but shifted by $6\pm3$~keV relative to Ref.~\cite{tang2010}. %
Apart from this small shift, our spectrum is consistent 
with both previous spectra as the level of statistics is insufficient 
to reveal the small discrepancies in the region around 
$E_{\alpha}=1.0$~MeV. %
Thus, our analysis shows that the spectrum of Ref.~\cite{azuma1994} 
is both supported by the better fit quality and in better agreement with 
the energy calibration of the present spectrum. %

Sub-Coulomb $\alpha$-transfer reactions provide an 
alternative route to determining $P_1 \gamma_{11}^2$ by constraining 
the ANC of the 7.12-MeV level which is related to $\gamma_{11}$ via 
Eq.~(44) in Ref.~\cite{deBoer2017}. %
Adopting the most recent and most precise ANC value of 
$(4.39\pm 0.59)\times 10^{28}$~fm$^{-1}$~\cite{avila2015} 
and assuming the channel radius to be $6.32\pm0.27$~fm (the 68.3\% 
confidence interval determined from the $\beta$-decay data, see 
the figure in Supplemental Material.), we obtain 
$P_1\gamma_{11}^2 = 4.44\pm 0.70\; \mu$eV in good agreement with Eq.~(\ref{eq:gamma}). %
The weighted average of the two is $4.71\pm 0.56\; \mu$eV, when statistical and systematic 
uncertainties are combined in quadrature, yielding a relative uncertainty of 12\%. %
We note that the less precise ANCs obtained in three previous 
$\alpha$-transfer studies are in good agreement with that of Ref.~\cite{avila2015}.

In conclusion, we have obtained the 
first accurate normalization of the 
$\beta$-delayed $\alpha$ spectrum of 
$^{16}$N and resolved a significant 
discrepancy between two previous high-precision 
measurements of the spectral shape. %
The branching ratio for $\beta$-delayed 
$\alpha$ emission is found to be 33\% larger than 
previously held and the value of $P_1\gamma_{11}^2$ 
inferred from the $\beta\alpha$ 
spectrum is increased by the same factor. %
Our value for $P_1\gamma_{11}^2$ is in good 
agreement with the value inferred from sub-Coulomb $\alpha$ 
transfer studies and has comparable precision. 
The weighted average of the two has an uncertainty of 12\%. %
Since the dominant term in the expression for the $E1$ 
capture cross section is proportional to $P_1\gamma_{11}^2$, 
our result implies that indirect measurements alone now 
constrain the $E1$ capture cross section to within close to 12\%, 
a remarkable result considering the large 
variability in the $S_{E1}(0.3)$ values reported over the 
last 60 years (Table~IV of Ref.~\cite{deBoer2017}). %
By further including direct measurements of the capture 
cross section as well as $\alpha$-scattering data it 
may be possible to reduce the uncertainty 
even further. %
Considering the progress made in recent 
years in constraining the other 
components of the $^{12}\text{C}(\alpha,\gamma)$ 
cross section, it may finally be possible to bring the 
uncertainty on the total cross section at 0.3~MeV 
below 10\%.

\paragraph{Acknowledgements}

\begin{acknowledgments}
We are grateful to the ISOLDE technical staff 
for providing excellent running conditions 
during the experiment, %
and thank the anonymous reviewers for their valuable 
comments and suggestions to improve the quality of the 
manuscript. %
This work has been supported by the European Research
Council under the ERC starting grant LOBENA, No.~307447, %
the Horizon 2020 Research and Innovation Programme under grant agreement No.~654002, %
the Spanish MINECO through projects FPA2015-64969-P, FPA2015-65035-P, and FPA2017-87568-P, %
the Romanian IFA grant CERN/ISOLDE, %
the United Kingdom Science and Technology Facilities Council, %
the FWO-Vlaanderen (Belgium) and GOA/2010/010 (BOF KU Leuven), 
the German BMBF under contract 05P15PKCIA (ISOLDE) and Verbundprojekt 05P2015. %
BJ acknowledges support from The Royal Society of Arts and Sciences in Gothenburg 
and OSK from the Villum Foundation through project no.\ 10117. 

\end{acknowledgments}

\section*{Supplemental Material}

The thickness of the catcher foil was determined from the energy loss 
of $\alpha$ particles from a standard spectroscopy source. The thickness 
was found to be $33\pm 3$~$\mu$g/cm$^2$ before the run and $36\pm 3$~$\mu$g/cm$^2$ 
after the run, indicating negligible changes in foil properties during the experiment. %
The two peaks in the $\beta\alpha$ spectrum of $^{18}$N were fitted with a Gaussian 
function, representing the experimental resolution, convoluted with the $\beta\nu$-recoil 
broadening function appropriate for a pure GT transition with the spin sequence 
$1^-\rightarrow1^-\rightarrow0^+$~\cite{clifford1989}. This broadening function has 
the approximate shape $f(x) \simeq 1 - 1.23x^2 + 0.23x^4$, where $x$ is the deviation from 
the mean $\alpha$-particle energy expressed as a fraction of the maximum deviation (36.3~keV 
for the 1081-keV peak and 38.3~keV for the 1409-keV peak). %
The experimental resolutions quoted in the Letter (30~keV for the two 60-$\mu$m DSSDs 
and 70~keV for the 40-$\mu$m DSSD) refer to the full width at half maximum (FWHM) of the Gaussian function. %
For the $\beta\alpha$ decay of $^{16}$N, the $\beta\nu$-recoil broadening function was 
approximated by a Gaussian with a width of 15~keV FWHM~\cite{azuma1994}. This width 
was added in quadrature with the experimental resolution to obtain the full Gaussian 
resolution for the $R$-matrix fit. %

We use the $R$-matrix parametrization of Ref.~\cite{brune2002} in 
which the spectrum is calculated as the incoherent sum of $p$-wave ($\ell=1$)
and $f$-wave ($\ell=3$) components given by,
\begin{equation}
N_{\ell} = N_{\alpha} \, f_{\beta} \, P_{\ell} \, \vert
\sum_{\lambda\mu} \widetilde{B}_{\lambda} \, 
\widetilde{\gamma}_{\lambda} \,
\widetilde{A}_{\lambda\mu}
\vert ^2 \; ,
\end{equation}
where $N_{\alpha}$ is the number of observed $\alpha$ particles, 
$f_{\beta}$ is the $\beta$-decay phase-space factor, $P_{\ell}$ is 
the penetration factor, $\widetilde{B}_{\lambda}$ is the feeding amplitude, 
$\widetilde{\gamma}_{\lambda}$ is the reduced $\alpha$ width, 
$\widetilde{A}_{\lambda\mu}$ is the level matrix, and the summation 
runs over the levels in the model. %
For bound levels, the feeding amplitude is given by~\cite{barker1988},
\begin{equation}\label{eq:feeding}
    \widetilde{B}_{\lambda}^{2} = \frac{b_{\beta,\lambda}}{\pi \, b_{\beta\alpha} \, f_{\beta,\lambda}}  
    \left( 1 + \widetilde{\gamma}_{\lambda}^2 \frac{dS_{\ell}}{dE} \right) \; ,
\end{equation}
where $b_{\beta,\lambda}$ is the branching ratio to the level in question, $b_{\beta\alpha}$ is the 
branching ratio for delayed $\alpha$ decay, and $S_{\ell}$ is the shift factor. %
Thus, to keep the product $\widetilde{B}_{\lambda}\widetilde{\gamma}_{\lambda}$ constant 
and thereby preserve the spectral shape, we must have $\gamma_{\lambda}^2 \propto b_{\beta\alpha} / b_{\beta,\lambda}$. %
The best-fit parameters for the spectrum of Ref.~\cite{azuma1994} are given in Table~\ref{tb:rmat} where 
$\widetilde{E}$ is the level energy relative to the $\alpha+{}^{12}$C threshold of 7161.92~keV. %
\newcolumntype{d}[1]{D{.}{.}{#1}}
\newcommand\mc[1]{\multicolumn{1}{c}{#1}}
\renewcommand{\arraystretch}{1.3}
\begin{table}[h]
\small
\centering
\caption{Best-fit $R$-matrix parameters for the spectrum of Ref.~\cite{azuma1994}; 
fit quality: $\chi^2/N = 94.3/79 = 1.19$; channel radius: 6.35~fm. Parameters in 
brackets were held fixed.}
\label{tb:rmat}
\begin{tabular}{c *{2}{d{2.4}} *{4}{d{1.4}}}
 \mc{Level ($n\ell$)} & \mc{$\widetilde{E}$ (MeV)} & \mc{$\widetilde{B}$} & \mc{$\widetilde{\gamma}$ (MeV$^{1/2}$)} \\
\hline
11 & [-0.0451] & [1.074] & 0.113 \\ 
21 & 2.388 & 0.394 & 0.488 \\ 
31 & [7.999] & -0.352 & 1.418 \\ 
13 & [-1.032] & [2.174] & [0.0626] \\ 
23 & [4.242] & -0.353 & [0.520] \\ 
\hline
\end{tabular}
\end{table}

As mentioned in the Letter, the $R$-matrix distribution obtained from 
the fit to the spectrum of Ref.~\cite{tang2010} is slightly wider and shifted by $-6$~keV 
relative to that obtained from the spectrum of Ref.~\cite{azuma1994}. %
We have confirmed this result using the best-fit parameters given in the respective 
papers~\cite{azuma1994, tang2010}. %
On the other hand, Ref.~\cite{deBoer2017} determines the shift to be only $-5-(-3.75)=-2.25$~keV 
(their TABLE~IX). While it is difficult for us to pinpoint the reason for this deviation with complete
certainty, we note that the 1-sigma resolutions adopted by the authors of Ref.~\cite{deBoer2017} 
for the $\beta\alpha$ spectra of Refs.~\cite{azuma1994, tang2010} were in fact FWHM resolutions, implying that the 
resolution was greatly overestimated in their fits. (The authors 
of Ref.~\cite{deBoer2017} have acknowledged this error to us in private communication.)

\begin{figure}
    \includegraphics[width=0.99\columnwidth,clip=true,trim=0 0 0 0]{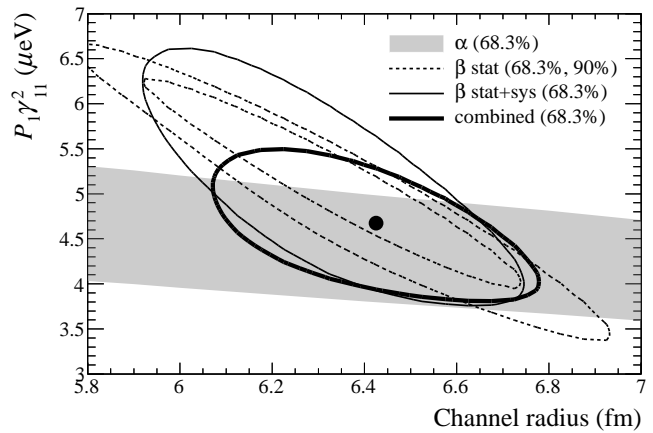}
    \caption{\label{fig:Pg2}Joint confidence region for $P_1\gamma^2$ and the 
    channel radius obtained from the $\alpha$-transfer study of Ref.~\cite{avila2015} (gray band) and 
    the present re-analysis of the $\beta\alpha$ spectrum of Ref.~\cite{azuma1994} 
    (thin contours). The combined confidence region is shown by the 
    thick contour and the preferred value is shown by the dot.}
\end{figure}

In Fig.~\ref{fig:Pg2} we compare the constraints imposed on $P_1\gamma_{11}^2$ 
by the present work to the constraints imposed by the most recent and precise $\alpha$-transfer 
study~\cite{avila2015}. For the $\beta$-decay data, statistical and systematic uncertainties 
were added in quadrature. The combined confidence region was determined with equal weights 
assigned to the two data sets. %

\end{document}